\documentclass{optica-article}
\makeatletter
\def\@journalname{}
\makeatother
\articletype{Research Article}

\usepackage{amsmath}
\usepackage{booktabs}
\usepackage{multirow}
\usepackage{graphicx}
\usepackage{subfig}
\usepackage{siunitx}
\usepackage{url}    

\newcommand{\Dr}{D/r_0}                    
\newcommand{\dtot}{d_{\mathrm{tot}}}       
\newcommand{\gphot}{g}                     
\newcommand{\Ameta}{A_{\mathrm{meta}}}     
\newcommand{\Ap}{A_{\mathrm{p}}}           
\newcommand{\nband}{n_{\mathrm{band}}}     

\begin{document}

\title{Anisoplanatic Blind Deconvolution using Phase Diversity}

\author{Shubhankar Jape,\authormark{1} Seungman Choi,\authormark{1} Peter Menart,\authormark{1} Andrew Schramka,\authormark{1} Pratyush Uppuluri,\authormark{2} Vaneet Aggarwal,\authormark{2} and Zubin Jacob\authormark{1,*}}

\address{\authormark{1}Elmore Family School of Electrical and Computer Engineering, Birck Nanotechnology Center, Purdue University, West Lafayette, IN 47907, USA.\\
\authormark{2}Edwardson School of Industrial Engineering, 315 Grant St, West Lafayette, IN 47906}

\email{\authormark{*}zjacob@purdue.edu}

\begin{abstract*}
Atmospheric turbulence distorts the wavefront of light from a distant scene and blurs its image. Over a field of view wider than the isoplanatic angle,
the distortion varies with direction and evolves in time, so no single wavefront describes the blur. Multiframe
blind deconvolution (MFBD) can recover the scene from measurements spread over a long acquisition, but falls short during fast imaging because of limited diversity across measurements. We present coded anisoplanatic blind deconvolution (CA-BD), in which a pupil-plane spatial light modulator (SLM) applies a different known phase code to every measurement, with no wavefront sensor or deformable mirror, so that the measurements carry diversity even within a short acquisition. A per-pixel single-layer estimator recovers the object and spatially varying
phases from the coded measurements. In simulation, a Cram\'er-Rao bound analysis shows that under anisoplanatism
direct imaging needs about 41 times the photons of coded imaging for equal phase precision in strong turbulence. At the stronger of two
turbulence strengths, CA-BD outperforms tuned MFBD on average at every burst length. It reduces the mean squared error $3.5\times$
for stationary turbulence and 2-3$\times$ for time-varying turbulence. Measurements taken within a few coherence times benefit most from the known coded SLM patterns, making faster SLMs the key to its full advantage in time-varying turbulence.
\end{abstract*}

\section{Introduction}
Long-range imaging, both on-sky and along the ground, is limited by atmospheric turbulence rather than by the optics
of the imaging system~\cite{choi2025photonstarvedimagingturbulencediffraction}. Refractive-index fluctuations along
the propagation path distort the wavefront of the incoming light~\cite{andrews2005laser}, and unless this distortion
is estimated and removed, the image stays far from the diffraction limit~\cite{mao2020imagereconstructionstaticdynamic}.
Two properties of turbulence, its variation in space and in time, make this estimation difficult. First, light from different points of the scene crosses
different parts of the turbulence, so the distortion varies across the field of view, a property known as
anisoplanatism~\cite{fried1982}. Second, the turbulence evolves in time, so measurements recorded at different instants see
different distortions. Retrieving the object together with turbulence that varies in both space and time is therefore
challenging for conventional imaging systems.

Existing methods for turbulence estimation and correction, such as speckle imaging~\cite{labeyrie1970speckle},
conventional adaptive optics (AO)~\cite{AO}, multiframe blind deconvolution (MFBD)~\cite{mfbd}, phase
diversity~\cite{gonsalves1982,paxman1992} and wavefront shaping (WS)~\cite{gfws,neuws}, assume a spatially invariant
wavefront. This assumption breaks down once the field of view exceeds the isoplanatic angle. To handle anisoplanatism,
these methods have been extended by spatially varying MFBD such as torchmfbd~\cite{torchmfbd}, by multiconjugate
adaptive optics (MCAO)~\cite{rigaut2020multiconjugateadaptiveopticsastronomy} and by machine-learning
methods~\cite{refractive_index,mao2020imagereconstructionstaticdynamic,zhang2025learningphasedistortionselective}.
MCAO relies on several wavefront sensors and deformable mirrors (DMs), which increases the hardware complexity and its
cost. Machine-learning methods depend on the data they were trained on and can fail when the turbulence departs from
the training data, for instance when the turbulence is stronger. Spatially varying MFBD models point spread functions (PSFs) that change
across the field, but apart from a fixed known aberration, usually defocus, in an optional phase-diversity channel~\cite{vannoort2005,torchmfbd}, its only source of diversity is the turbulence itself. When the turbulence evolves slowly or the acquisition window is short, consecutive measurements differ little, and spatially varying MFBD struggles to estimate the turbulence and the object accurately.

We propose coded anisoplanatic blind deconvolution (CA-BD), a hardware--software co-design that estimates spatiotemporally varying turbulence where these methods
break down. Building on our coded detection for static, isoplanatic
turbulence (TAP-BD)~\cite{choi2025photonstarvedimagingturbulencediffraction}, a spatial light modulator (SLM) in the pupil
plane adds a different known phase code to every measurement, which supplies the diversity that the turbulence alone does not provide within a short acquisition window. From the coded measurements we estimate a common object and a single turbulent layer that covers the whole field of view. Each part of the field sees this layer through its own shifted footprint (Fig.~\ref{fig:forward}(a)), so the one layer gives the spatially varying phases. The layer is a per-pixel phase screen, not restricted to a modal expansion such as Zernike polynomials, and is optimized by the gradient-based Adam
method~\cite{kingma2015adam}, with the object obtained in closed form at each step. We show in simulation that coded
measurements recover the spatially varying phases and the object for static and time-varying turbulence, and that they
reach higher image quality than MFBD on uncoded measurements at the same photon budget. We also show that a current 1.5\,kHz SLM already codes every measurement on bursts that span several coherence times, on which CA-BD still leads MFBD on average. The SLM refresh rate limits coded acquisition on shorter bursts, so faster SLMs would move coded imaging toward real-time turbulence correction.

\section{Forward Model and Estimation}

\begin{figure*}[t]
  \centering
  \includegraphics[width=0.95\textwidth]{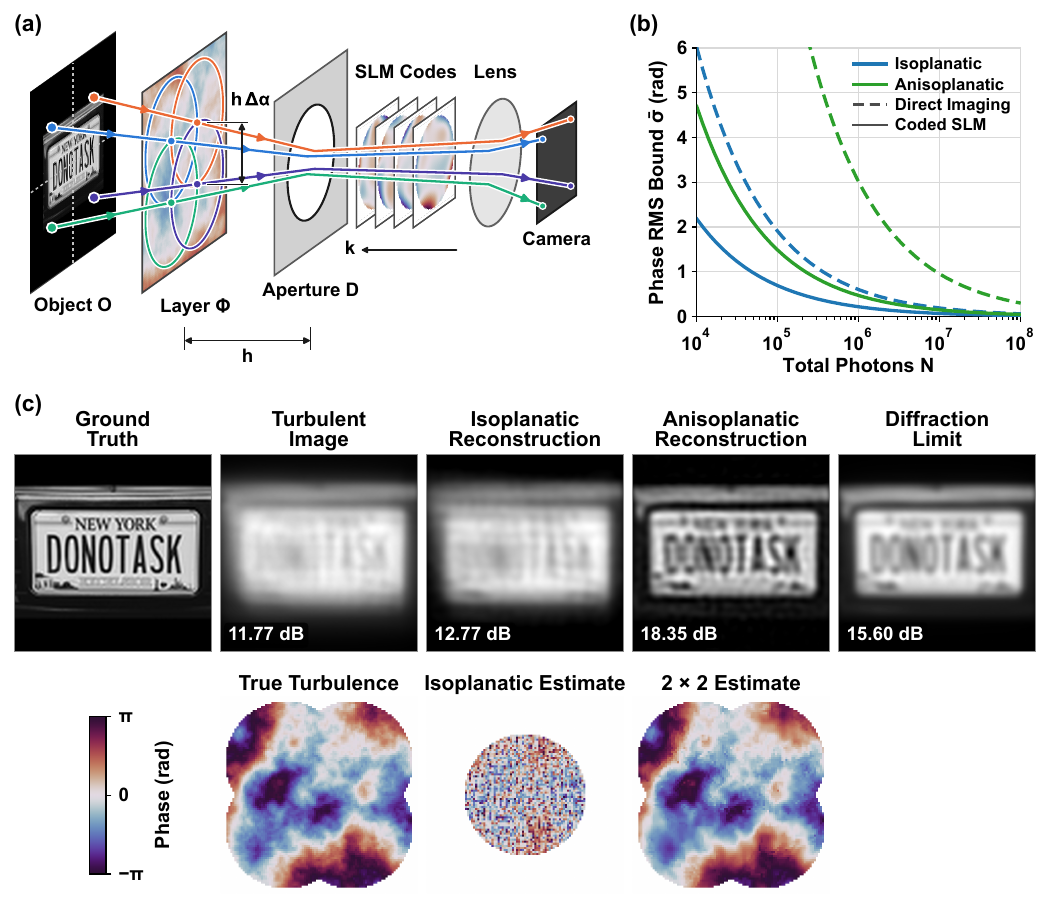}
\caption{Anisoplanatic imaging with a coded spatial light modulator (SLM).
(a)~Optical layout: Each $2\times2$ object block sees the turbulent layer $\Phi$ at distance $h$ through its own
footprint. Adjacent footprints are offset by $h\,\Delta\alpha$, half the aperture diameter ($D/2$) in the simulations, so neighboring footprints overlap. A
pupil-plane SLM adds a known code $\theta_k$ to measurement $k$.
(b)~Phase-only Cram\'er-Rao bound: The bound is plotted versus the total
photon number $N$ for isoplanatic (blue) and anisoplanatic (green) turbulence, recorded by direct imaging (dashed) or
through the coded SLM (solid). Phase diversity lowers the bound in both cases.
(c)~Isoplanatic versus anisoplanatic reconstruction: The top row shows the images, with
labels giving the registered PSNR, and the bottom row shows the true and estimated phases.}
  \label{fig:forward}
\end{figure*}

We begin by describing the coded anisoplanatic imaging system and its forward model (Fig.~\ref{fig:forward}(a)).
Light from an incoherent object $O$ crosses a single turbulent layer
$\Phi$ at distance $h$. Each part of the field, or block, sees the layer through its own footprint. An SLM conjugate
to the pupil then adds a known phase code $\theta_k$ to measurement $k$, so the $K$ recorded measurements are differently coded
images of one object through one layer. Appendix~\ref{app:sim} gives the simulation details used to generate the von K\'arm\'an phase screen.

\subsection{Anisoplanatic coded image formation}\label{sec:formation}

Atmospheric turbulence is distributed continuously along the propagation path, with a strength profile
$C_n^2(z)$~\cite{roddier1981,andrews2005laser}. Imaging and propagation models commonly replace this volume by thin
phase screens, each collecting the phase of one slab of turbulence~\cite{roddier1981,schmidt2010numerical}. We
approximate it by a single such screen, the simplest model that produces anisoplanatism: a von K\'arm\'an phase layer
$\Phi$ at distance $h$ from a circular aperture of diameter $D$, with outer scale $L_0 = 40D$~\cite{andrews2005laser}.
The screen is generated by the standard fast-Fourier-transform (FFT) method with
subharmonics~\cite{lane1992screen,schmidt2010numerical}. Each screen is then rescaled to a nominal Fried parameter
$r_0$. Throughout the paper, $\Dr$ denotes this nominal value: the effective strength of a
screen is higher on average and varies between realizations, and only the screens marked actual
skip the rescaling.

Anisoplanatism for one layer is characterized by the isoplanatic angle $\theta_0 = 0.314\,r_0/h$~\cite{fried1982}. In
practice $r_0$ and $\theta_0$ are measured routinely by multi-aperture scintillation sensor and differential image
motion monitor (MASS--DIMM) instruments~\cite{kornilov2007}. The measured pair then gives the layer height directly,
$h = 0.314\,r_0/\theta_0$. This height sets the footprint shifts of the forward model, Eq.~(\ref{eq:shift}), and
$\theta_0$ sets how many blocks the field needs. In a real atmosphere the turbulence is spread over height, and
this $h$ is the effective height at which one screen reproduces both $r_0$ and $\theta_0$ of the
profile~\cite{fried1982}.

In a block model, the field of view is divided into $B$ blocks, each imaged with the PSF of its center direction, and the windows of Eq.~(\ref{eq:forward}) blend these PSFs across the block borders. A block model faithful to this layer would need blocks about as wide as $\theta_0$. In the simulated geometry, this
means about 6.5, 13 and 26 blocks per side at $\Dr = 2$, 4 and 8. Each block has
its own PSF in every measurement, so with $B$ blocks and $K$ measurements the estimator of Section~\ref{sec:estimator} evaluates $BK$ PSFs per
iteration. The unregularized form of its object step also sets a conservative design rule of more coded measurements than
blocks, $K > B$. Under this rule, a $13\times13$ partition at $\Dr = 4$ would call for more than
169 measurements. Each iteration would then need at least 28,730 PSF evaluations, more than 100 times the 256 of a
$2\times2$ model with 64 measurements. As a proof of concept, we use $B = 4$ blocks in a $2\times2$ partition, indexed $b = 0, \dots, B-1$, so that four representative PSFs, interpolated by the windows, describe the field. A finer partition would use the same estimator at the cost above. 

Each block is treated as a distant source, so it sees the layer through a footprint of diameter $D$. Together, the footprints of all blocks cover a larger region of the layer, the metapupil. The footprint
is displaced on the layer by the metapupil shift
\begin{equation}
  s_b = h \tan \alpha_b,
  \label{eq:shift}
\end{equation}
where $\alpha_b$ is the field angle of the block center. The pupil phase $\phi_b$ of block $b$ is the crop of $\Phi$
at $s_b$ with its aperture piston removed. The shifts follow from Eq.~(\ref{eq:shift}) and are smaller than the footprint diameter $D$, so the footprints
of neighboring blocks overlap (Fig.~\ref{fig:forward}(a)) and their phases must agree on the shared pupil pixels.

We render the data from simulated skies in which the field-dependent turbulence itself follows the block model, $2\times2$
unless stated otherwise. The results therefore describe such skies rather than a physical single-layer sky over the
same field. The single-layer picture also requires the seeing blur $\lambda/r_0$ at wavelength $\lambda$ to stay
below $\theta_0$, which in the simulated geometry holds up to about $\Dr = 4.4$. The
strengths $\Dr = 6$--10 of the measurement-count sweep (Fig.~\ref{fig:framecount}) lie beyond this limit.

The SLM code $\theta_k$ of frame $k = 1, \dots, K$ is a random combination of the 20 lowest-order Zernike modes above
piston, drawn anew for every frame and shared by all blocks. As in phase diversity~\cite{gonsalves1982}, but with a known aberration that changes with every measurement, the codes supply the diversity that can make the phase estimable from intensity alone under the band-limited object model of Appendix~\ref{app:fisher}.

In measurement $k$, block $b$ has the pupil phase $\phi_{b,k}$, equal to $\phi_b$ for a frozen screen, and the PSF
\begin{equation}
  H_{b,k} \propto \bigl| \mathcal{F}\{ P \exp( i [ \phi_{b,k} + \theta_k ] ) \} \bigr|^{2},
  \label{eq:psf}
\end{equation}
normalized to unit sum over the image grid. Here $P$ is the pupil function of the circular aperture, and
$\mathcal{F}$ is the two-dimensional discrete Fourier transform. The noise-free image is the composite
\begin{equation}
  y_k = \sum_{b=0}^{B-1} (O w_b) * H_{b,k},
  \label{eq:forward}
\end{equation}
where $*$ is circular convolution on the image grid and the $w_b$ are raised-cosine windows that sum to one at
every pixel. This is the standard partition-of-unity, or efficient-filter-flow, model of spatially varying
blur~\cite{nagy1998,hirsch2010}. Because the windows blend PSFs rather than phases, the model only approximates a
continuously varying PSF, least well at the block junctions, where the object sits.

The total photon budget of $1\times10^{9}$ detected photons is split evenly over the $K$ measurements, and detection is
photon counting,
\begin{equation}
  n_k \sim \mathrm{Poisson}(\gphot \, y_k), \qquad
  \gphot = 10^{9} \Big/ \Bigl( K \sum_r O(r) \Bigr).
  \label{eq:poisson}
\end{equation}
Every reconstruction uses $K = 64$ measurements unless $K$ is varied.

The forward model rests on five assumptions (Appendix~\ref{app:sim}): a thin layer seen from distant sources in monochromatic light; skies that follow the block model; an ideal modulator with exactly known codes, instantaneous switching and instantaneous exposures; Poisson noise; and a screen that is frozen except in the time-varying studies of Section~3.

We first ask what the codes add to the data before any estimator is run. The Fisher information of the Poisson
likelihood of Eq.~(\ref{eq:poisson}) measures how strongly the measurements constrain the object and the
phase~\cite{chao2016fisher}. Its inverse, the Cram\'er--Rao bound (CRB), bounds the error variance of any unbiased estimator of the object and the phase, and it has been used to analyze phase-diverse wavefront sensing~\cite{lee1999crb}.
Fig.~\ref{fig:forward}(b) plots the square root of the aperture-averaged phase-only CRB against the total photon
number summed over $K = 10$ measurements at $\Dr = 20$. The object is taken as known, so that the phase is isolated
and direct imaging has a finite CRB. We group the four cases in the plot into two categories, by turbulence and by
optical setup. The two turbulence regimes are isoplanatic and anisoplanatic, and the two optical setups are direct
imaging, without an SLM, and imaging through the coded SLM. All four curves fall as the inverse square root of the
photon number, as expected for Poisson data. In both isoplanatic and anisoplanatic turbulence, the coded-SLM bound
remains below the direct-imaging bound. This gives intuition as to why a coded SLM helps turbulence estimation.
Furthermore, the gain of the coded SLM over direct imaging is larger for anisoplanatic turbulence.
In the reduced geometry of Appendix~\ref{app:fisher}, direct imaging needs about 7.6 times as many photons as coded
measurements to reach the same CRB when the turbulence is isoplanatic, whereas it needs about 41 times as many when
it is anisoplanatic. Each known code changes how the phase maps to intensity, the mechanism of phase
diversity~\cite{gonsalves1982}, so each coded measurement adds a new view of the phase, whereas frozen uncoded
measurements add only photons. For isoplanatic turbulence, our results agree with the lower CRB reported for coded
detection in Ref.~\cite{choi2025photonstarvedimagingturbulencediffraction}. Appendix~\ref{app:fisher} gives both
computations and the identifiability analysis.

\subsection{Temporal evolution}\label{sec:temporal}

We model the time evolution with Taylor's frozen-flow hypothesis~\cite{taylor1938}: one screen slides across the pupil at constant speed $v$ without changing shape.
Only the time-varying studies of Section~3 and the analyses built on them use this drift. The temporal scale is the
coherence time $\tau_0 = 0.314\,r_0/v$, and we express drift by $\dtot$, the number of coherence times an
acquisition of $K$ measurements spans. Measurement $k$ sees the screen displaced by $(k-1)\,\delta$, with per-measurement shift
\begin{equation}
  \delta = \dtot \, v\tau_0 / K,
  \label{eq:delta}
\end{equation}
where $v\tau_0 = 0.314\,r_0$ is the screen travel in one coherence time.

\subsection{Per-pixel estimator}\label{sec:estimator}

We estimate the phase as a pixel screen, in which every pixel value is a separate unknown, rather than as a modal expansion. A pixel screen can represent phase at every scale the grid samples, so the estimator has no modal-truncation floor and its only representation limit is the block model. With one layer, the unknowns are the object and the layer's phase screen. Under drift, the estimator splits the $K$ measurements into $E$ consecutive groups, called epochs, and treats the screen as frozen within each: it estimates one screen $\hat\Phi_e$ per epoch $e = 0, \dots, E-1$ and models measurement $k$ with the screen of its epoch $e(k)$. A frozen screen has $E = 1$. The estimator is told which measurements form each epoch, but not how the screen moves between them, so $E$ is an estimator choice rather than a property of the data. The block phases $\hat\phi_{b,e}$ are crops of $\hat\Phi_e$ at the shifts of
Eq.~(\ref{eq:shift}), so the estimator can represent a set of block phases only if it comes from one screen. It is
also given the exact codes and the true shifts, and hence the layer distance, which Section~4 discusses calibrating
from the data.

From the measurements, the codes and the shifts, CA-BD obtains the screens by minimizing over $\hat\Phi$ the object-profiled loss,
\begin{equation}
  J(\hat\Phi) = \min_{O} \; \sum_{k=1}^{K} \Bigl\| n_k - \gphot \sum_{b=0}^{B-1}
  (O w_b) * H_{b,k}(\hat\phi_{b,e(k)}) \Bigr\|^2 + \mu \, T(\hat\Phi),
  \label{eq:objective}
\end{equation}
where $\mu$ weights an anisotropic total-variation penalty $T$ on the screens~\cite{rudin1992tv}. The
data term is least squares although the noise is Poisson, a mismatch that is mild at the photon levels used here.
For fixed screens, a closed-form object step approximates the inner minimum over $O$. Because convolution becomes multiplication in the Fourier domain, the step fits each spatial frequency separately: a small regularized least-squares solve matches that frequency of the $B$ windowed object pieces $O w_b$ to the same frequency of the $K$ measurements, and the pieces are then merged into one object and refined (Appendix~\ref{app:est}). The phase gradient is taken by automatic differentiation through this step. Because this
step does not reach the exact inner minimum, the estimator is not exact variable
projection~\cite{golub1973separable}, and its gradient is exact for the approximated loss rather than for $J$.
With $B = 1$ the same estimator is isoplanatic, and with identical settings it gives the isoplanatic reconstruction
of Fig.~\ref{fig:forward}(c).

The screens were updated by Adam~\cite{kingma2015adam} from a single zero-phase start, with fixed settings
that depend only on the $\Dr$. The reported iterate is selected by the value of Eq.~(\ref{eq:objective})
alone, with no ground-truth quantity. 
We score each reconstruction by the registered peak signal-to-noise ratio (PSNR) of the image and by the root-mean-square (RMS) error of the estimated block phases.

While the coded SLM lowers the bound in both regimes, we now ask whether anisoplanatic turbulence, which is more
realistic for wide fields, also needs an anisoplanatic model in the per-pixel estimator (Fig.~\ref{fig:forward}(c)).
We reconstruct the same 64 coded measurements with two models. The measurements are rendered at $\Dr = 2$ by dividing
the field of view into four quadrants, the $2\times2$ blocks of Section~\ref{sec:formation}, each with its own PSF, blended at the
seams by the windows of Eq.~(\ref{eq:forward}). The first model is isoplanatic, the $B = 1$ case of CA-BD, and the
second is the matched $2\times2$ model. The $2\times2$ reconstruction reaches a registered PSNR of 18.35\,dB, above
the 15.60\,dB of the diffraction-limited image, which is a single unprocessed exposure, not an upper bound. In
contrast, the isoplanatic reconstruction reaches only 12.77\,dB, close to the 11.77\,dB of the turbulent image. The
phases explain the gap. The $2\times2$ model recovers the block phases to an RMS error of 0.12\,rad, whereas the
isoplanatic phase lies 1.30\,rad from them on average, farther than a zero-phase estimate (1.02\,rad). Although the
footprints of the four quadrants overlap, their block phases differ from one another by 1.1 to 1.4\,rad RMS. The
single phase absorbs this field dependence as a non-physical wavefront, a fine speckle rather than a smooth screen.
Supplement~1, Sec.~S1, repeats the comparison for two screen realizations at $\Dr = 2$ and 4.

\begin{figure*}[t]
  \centering
  \includegraphics[width=1.0\textwidth]{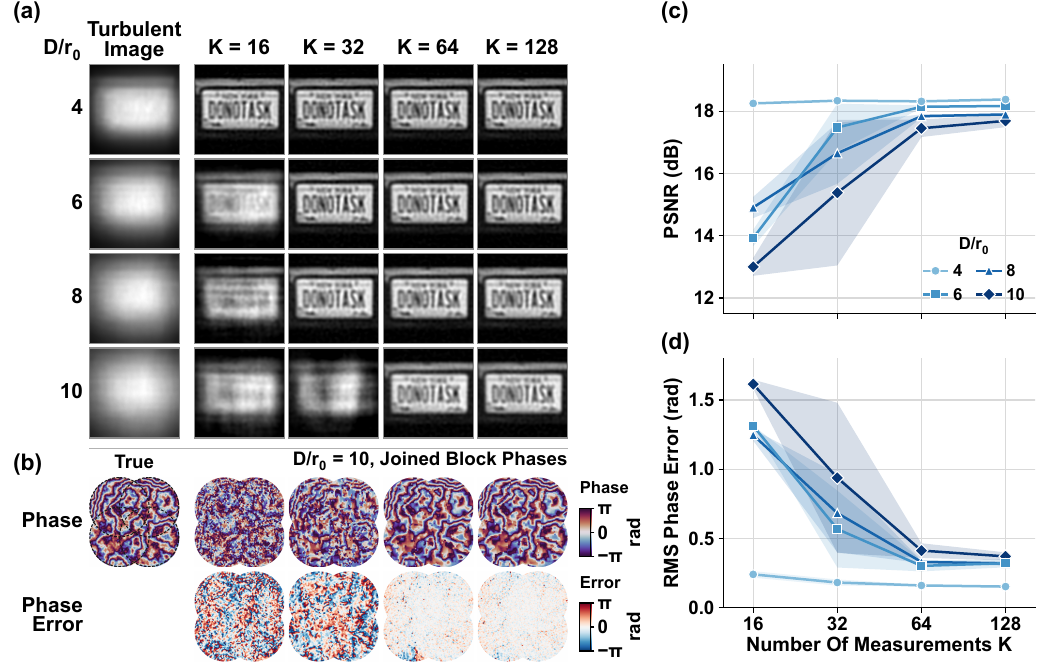}
 \caption{CA-BD reconstruction versus the number of coded measurements at a fixed photon budget, averaged over two
screen realizations.
(a)~Example reconstructions from one realization (turbulence strength increasing by row): The left column shows the
turbulent image, and the other columns show reconstructions from increasing numbers of coded measurements.
(b)~Joined phase of the four blocks in the strongest turbulence: The four overlapping block footprints, outlined on the true phase, form one phase map. The top row shows the true phase and its estimate for each
measurement count, and the bottom row shows the estimation error.
(c,~d)~Registered PSNR (c) and RMS phase error (d) versus the number of measurements: Darker blue marks stronger
turbulence. Lines give the mean over the two realizations, and shaded bands their range.}
  \label{fig:framecount}
\end{figure*}
\section{Results}

We first ask how coded SLM measurements perform for static anisoplanatic turbulence. To understand reconstruction performance, we vary the number of coded measurements $K$ at a fixed total photon budget and at four turbulence strengths (Fig.~\ref{fig:framecount}). Each
measurement carries a different known code and so adds an independent view of the screen. The sweep shows two
trends. At every strength, more measurements improve the estimate on average (reconstructions in Fig.~\ref{fig:framecount}(a), averages in (c,~d)), even though each measurement then
receives fewer photons. The number of measurements needed to recover the phase and the object grows with the
turbulence strength. At $\Dr = 4$, $K = 16$ measurements already suffice: beyond that the registered PSNR stays flat (Fig.~\ref{fig:framecount}(c)) and only the phase error drops (Fig.~\ref{fig:framecount}(d)). At
this strength the residual phase is already small, so the image is limited by the object step rather than by the
wavefront. Extra measurements therefore improve the phase estimate without changing the
image. At $\Dr = 6$ to 10, more measurements are needed to 
improve both the phase and the image. Fig.~\ref{fig:framecount}(b) shows the estimated phase at $\Dr = 10$ and the error between the estimated phase and the ground truth. By $K = 64$ the mean phase error is below 0.5\,rad at $\Dr = 6$, 8 and 10 (Fig.~\ref{fig:framecount}(d)), and at $K = 128$ the registered PSNR recovers to $\sim$18\,dB at every strength in Fig.~\ref{fig:framecount} (17.7--18.4\,dB). A similar trade-off between
turbulence strength and the number of measurements was reported for isoplanatic coded imaging~\cite{choi2025photonstarvedimagingturbulencediffraction},
and here it persists with four block phases cropped from one screen. Stronger
turbulence puts more phase in each block, so there is more to recover. With few measurements,
several different turbulent phase screens can explain the coded images about equally well. Each added measurement, with its own coded phase,
rules out more of them, until only screens close to the true one, up to piston and tilt, fit all the codes at once, and the image sharpens as the phase
is found. This agrees with what has been reported previously for NeuWS~\cite{neuws} and TAP-BD~\cite{choi2025photonstarvedimagingturbulencediffraction}. More coded measurements at a fixed photon budget thus extend CA-BD to stronger turbulence.

\begin{figure*}[t]
  \centering
  \includegraphics[width=1\textwidth]{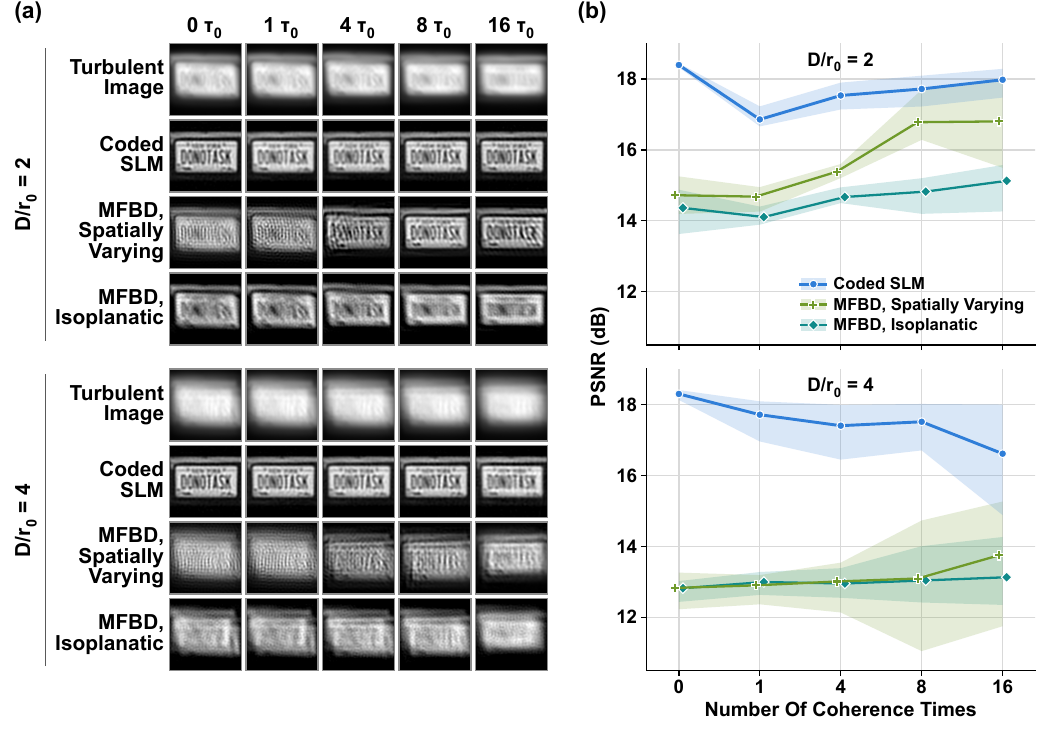}%
 \caption{Reconstruction through time-varying turbulence, averaged over three screen realizations.
(a)~Example reconstructions (weaker turbulence, top; stronger, bottom): Rows show the turbulent image, CA-BD (labeled
Coded SLM) and the two MFBD baselines, and columns give the number of coherence times the burst spans, from frozen (0)
to 16.
(b)~Registered PSNR versus the number of coherence times: CA-BD performs better on average than both MFBD baselines. Lines give the mean over the three realizations, and shaded bands their range.
All methods use $K = 64$ measurements and $1\times10^{9}$ detected photons per burst, and the baselines see uncoded
measurements.}
  \label{fig:timevarying}
\end{figure*}

While CA-BD performs well for static turbulence, to simulate a more realistic scenario, we now let the screen drift during the acquisition and compare CA-BD with two MFBD pipelines (Fig.~\ref{fig:timevarying}). We model this time-varying turbulence using Taylor's frozen-flow hypothesis (Section~\ref{sec:temporal}), fix each burst at $K = 64$ measurements and vary the number of coherence times it spans: 0 (frozen), 1, 4, 8 and 16. We ran three
screen realizations at $\Dr = 2$ and 4, each shared by all five drifts, so the points along each curve are paired.
Our estimator models each burst with one frozen epoch per coherence time spanned, or a single epoch for a frozen
screen. We compare CA-BD for time-varying turbulence with two MFBD baselines based on torchmfbd~\cite{torchmfbd}. Because torchmfbd does not accept a known pupil code, we compare acquisition-and-reconstruction pipelines at an equal photon budget, in which MFBD exploits the diversity of the turbulent phase itself for reconstruction. Both baselines
reconstruct uncoded measurements of the same scene, recorded through the same screen. The baseline
\emph{MFBD, isoplanatic} fits one wavefront per measurement over the whole field~\cite{mfbd,vannoort2005}, and
\emph{MFBD, spatially varying} is our implementation of the published spatially varying
model~\cite{torchmfbd}. With its maps and kernels set to our windows and block PSFs, this model reproduces
Eq.~(\ref{eq:forward}) term for term.

We simulate time-varying turbulence corresponding to two different turbulence strengths and report our reconstruction performance in Fig.~\ref{fig:timevarying}. In the case of frozen turbulence at $\Dr = 2$, CA-BD outperforms both baselines, because the baselines' uncoded measurements contain no diversity: they differ only in photon noise and so admit many possible reconstructions.
Under time-varying turbulence, when all measurements are captured within one coherence time, CA-BD falls because our estimator models the burst with a single frozen epoch, which must represent a drifting screen. As we increase the number of coherence times within our captured measurements, the coded reconstruction recovers to 17.5--18.0\,dB, because the diversity provided by multiple epochs improves the object reconstruction. As we increase the turbulence strength to $\Dr = 4$ the coded reconstruction falls as number of coherence times in the measurement increases. This happens because more epochs leave fewer measurements for each frozen screen and stronger turbulence needs more measurements per screen (Fig.~\ref{fig:framecount}). 

In contrast, time-varying turbulence helps the baselines at $\Dr = 2$. The spatially varying MFBD rises with an increase in number of coherence times, while the isoplanatic baseline changes little. The moving screen gives their uncoded measurements some of the diversity that CA-BD supplies by design, so the gap narrows. At 8 and 16 coherence times the coded band and that of the spatially
varying baseline overlap, but CA-BD still leads on average. Fig.~\ref{fig:timevarying}(a) shows the 8-coherence-time case at $\Dr = 2$, in which both reconstructions
resolve the fine detail of the object.
As we increase the turbulence strength to $\Dr = 4$, the diversity provided by the drifting turbulence does little for the baselines, which improve only slightly with the number of coherence times, while the coded reconstructions in Fig.~\ref{fig:timevarying}(a) resolve the fine detail throughout. On average, CA-BD leads the better baseline at each burst length by 2.9 to 5.5\,dB at $\Dr = 4$, against 0.9 to 3.7\,dB at $\Dr = 2$. At $\Dr = 4$, CA-BD and the baselines come closest at 16 coherence times, but CA-BD still averages higher. Overall, the coded lead is largest in the stronger turbulence, where drift adds little diversity to uncoded measurements.

\begin{figure*}[t]
  \centering
  \includegraphics[width=1.02\textwidth]{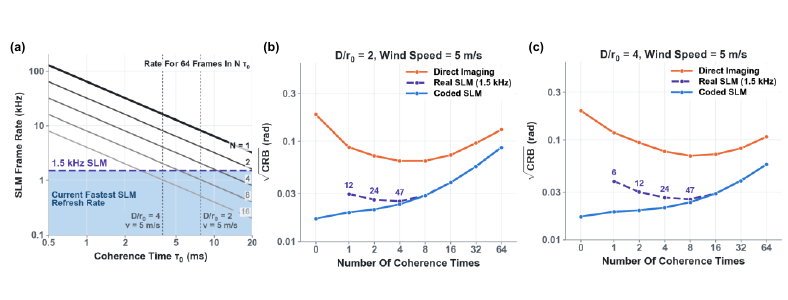}%
 \caption{Coded acquisition under drift at realistic SLM refresh rates.
(a)~Required frame rate versus coherence time $\tau_0$: Solid lines (black for $N = 1$) show the rate needed to record a burst of coded
measurements within $N$ coherence times. The dashed line marks a 1.5\,kHz SLM, and the shading marks the rates within its reach. Dotted lines mark $\tau_0$ for the two turbulence strengths of (b) and (c).
(b,~c)~Phase bound versus burst length (weaker turbulence in (b), stronger in (c)): The square root of
the CRB is plotted versus the number of coherence times the burst spans, for direct imaging (orange), an ideal coded
SLM with a new pattern on every measurement (blue), and a 1.5\,kHz SLM that must repeat patterns on short bursts
(purple dashed; numbers give its number of distinct patterns). On the computed bursts the 1.5\,kHz SLM reaches at best about
1.5 times the frozen-turbulence bound, and both coded curves stay below direct imaging.
All curves use the same number of measurements and photons.}
  \label{fig:rate}
\end{figure*}

With CA-BD outperforming both baselines on average at every burst length, we now ask which of these bursts a realistic SLM can record. Hence, we ask whether a real SLM can switch fast enough to record such bursts (Fig.~\ref{fig:rate}).
TAPBD~\cite{choi2025photonstarvedimagingturbulencediffraction} assumed turbulence that stayed static during acquisition and named faster SLMs and
cameras as one remedy. Recording $K = 64$ coded measurements within $N$ coherence times takes a camera running at
$K/(N\tau_0)$ (Fig.~\ref{fig:rate}(a)). Because every coded measurement carries its own pattern, the SLM must keep
the same pace, whereas direct imaging needs only the camera. A
5\,m/s crosswind gives $\tau_0 = 7.85$\,ms at $\Dr = 2$ and 3.93\,ms at $\Dr = 4$, so a one-coherence-time burst
needs about 8.2 and 16.3\,kHz. This wind speed lies in the middle of the 2--7\,m/s range of median near-ground wind speeds at astronomical sites, and the two values of $\tau_0$ bracket the 4.2--5.6\,ms median coherence times measured there~\cite{schock2009tmt}. We take 1.5\,kHz as the rate of a realistic SLM, close to the fastest settled rate
shown for a multilevel phase modulator~\cite{rocha2024mems}. Such an SLM is too slow for a one-coherence-time burst. It is also slower than high-speed cameras, which exceed 2\,kHz at full frame~\cite{ximea2026camera}, so it is the SLM rate, set against the coherence time, that limits coded acquisition. A 1.5\,kHz SLM needs 42.7\,ms to show 64 distinct patterns, which is $N^\ast = K/(f_{\mathrm{SLM}}\tau_0) \approx 5.4$ coherence times at $\Dr = 2$ and 10.9 at $\Dr = 4$, where $f_{\mathrm{SLM}}$ is the SLM rate. It can therefore give every measurement its own pattern on any burst of at least $N^\ast$ coherence times. The turbulence evolves during such bursts,
whereas on shorter ones the camera outruns the SLM and consecutive measurements share a pattern. In Fig.~\ref{fig:rate}(a), these bursts are the solid lines that cross the dotted $\tau_0$ lines inside the shading: 8 and 16 coherence times at $\Dr = 2$, and 16 at $\Dr = 4$. Shorter bursts, which need a faster SLM, give the ideal coded SLM of Fig.~\ref{fig:rate}(b,~c) a lower bound than these longer ones.

To compare a 1.5\,kHz SLM with an ideal coded SLM and with direct imaging, we computed the phase CRB, with the object known, for three acquisitions that share
the drifting screen, the frame times and the photon budget (Fig.~\ref{fig:rate}(b,~c)). They are direct imaging (the acquisition used by the MFBD baselines), an
ideal coded SLM that changes the pattern on every measurement, and the 1.5\,kHz SLM. The ideal coded SLM reaches its lowest $\sqrt{\mathrm{CRB}}$ for frozen turbulence, and its CRB rises with burst length as the drifting screen sweeps more unknown pixels through the pupil. To our knowledge, the phase CRB of direct imaging has not been computed before as a function of the number of coherence times a burst spans. As the burst spans more coherence times, the diversity due to the drifting turbulent phase improves estimation, and the bound decreases even as the number of pixels to be estimated increases. Beyond about 4 coherence times at $\Dr = 2$ and 8 at $\Dr = 4$, however, the bound increases as the number of pixels to be estimated goes up and the diversity provided is no longer as significant relative to the number of pixels. Under drift, the 1.5\,kHz SLM cannot reach the frozen-turbulence minimum of the ideal coded SLM. On bursts shorter than $N^\ast$ coherence times, the 1.5\,kHz SLM cannot apply 64 distinct patterns. The numbers above the purple line (Fig.~\ref{fig:rate}(b,~c)) correspond to the number of distinct patterns that can be applied at each of these burst lengths. The purple line demonstrates that, even with fewer distinct patterns, the $\sqrt{\mathrm{CRB}}$ of the 1.5\,kHz SLM is still lower than that of direct imaging.

A 1.5\,kHz SLM thus already codes every measurement on the longer bursts identified above, on which CA-BD leads both MFBD baselines on average (Fig.~\ref{fig:timevarying}), and its $\sqrt{\mathrm{CRB}}$ stays below that of direct imaging on every burst computed. On shorter bursts, the SLM rate is the limit of coded acquisition under drift. At the same photon budget, faster SLMs, with
a camera to match, would fit the coded measurements within a coherence time and bring the $\sqrt{\mathrm{CRB}}$ toward its frozen-turbulence value.

\section{Discussion}
Our forward model rests on the five assumptions of Section~\ref{sec:formation}, and the estimator is also given the layer distance.
The main hardware limitation is the modulator rate. Every reconstruction used a new, exactly known pattern on each
measurement, switched instantaneously. A realistic modulator repeats patterns on short bursts, and the longer bursts it can
fill span an evolving screen (Fig.~\ref{fig:rate}). Faster SLMs, with a camera to match, are the remedy. Measuring
the displayed codes and their switching in place, and reconstructing from repeated patterns, would test the remaining
modulator idealizations. Once the measurements can be coded within a coherence time, the same modulator could also apply
the estimated phase as a correction. TAPBD~\cite{choi2025photonstarvedimagingturbulencediffraction} has applied its reconstructed phase as an
optical correction for static, isoplanatic turbulence. In the pupil plane such a correction holds in the direction of
one block, whereas a corrector conjugate to the layer could correct across the field. Faster SLMs would thus let the
method work toward correcting turbulence in real time, provided the reconstruction, now far slower than a coherence
time, is accelerated to match.

A single frozen-flow layer restricts
the method to turbulence at one distance that moves as a rigid screen. A real profile calls for one estimated screen
per dominant layer, each sliding with its own flow, much as multiconjugate adaptive optics conjugates one corrector
to each layer~\cite{beckers1988}. A measured $\theta_0$~\cite{kornilov2007} and $r_0$ fix the distance of an
equivalent layer, and fitting the shift magnitude as one more unknown would calibrate it from the measurements. Each block
is also treated as a distant source, so a scene at finite range would need footprints and shifts computed for that
range. Our measurements follow the $2\times2$ block model, whose windows blend PSFs rather than phases, so the results
describe such measurements rather than one physical layer viewed across the same field. A finer estimator partition can follow turbulence with more isoplaantic patches, and the unregularized form of its object step
sets a conservative design rule of more frames than blocks, $K > B$. Light of a single
wavelength restricts the method to narrowband imaging, and a broadband scene would need a forward model that sums the
coded PSFs over the spectrum. The simulated measurements carry Poisson photon noise at a fixed budget, and at low flux a Poisson denoiser
could precede the estimator, as in the isoplanatic coded-SLM method of TAPBD. CA-BD performs
well in these simulations. The next steps are to demonstrate it in experiment and to extend it to several layers.
On the coded-SLM testbed of that work, a known phase plate away from the pupil plane would give each block its
own footprint and a ground-truth phase.

\section{Conclusion}
We have shown that a known random phase code on every measurement lets a per-pixel single-layer estimator
recover the object and the turbulent layer across an anisoplanatic field. On the same coded measurements at
$\Dr = 2$, the anisoplanatic model reaches a registered PSNR of $\sim$18.4\,dB, against $\sim$12.8\,dB for an
isoplanatic model. More measurements at a fixed photon budget improved anisoplanatic estimation accuracy. Under time-varying turbulence and given equal number of photons, CA-BD improves reconstruction performance compared to MFBD. This is due to the increase in diversity provided by known phases compard to diversity given by turbulence. In the simulated geometry, coding keeps the Cram\'er--Rao bound on the
phase below that of direct imaging at every burst length. This holds even for a realistic SLM, which already gives every measurement its own pattern on bursts that span several coherence times and repeats patterns only on shorter ones. The modulator rate is what limits coded acquisition on these shorter bursts. At the same photon budget, faster SLMs, with a camera to match, would bring the coded SLM toward its frozen-turbulence performance. Because the coding adds a
pupil-plane SLM to a conventional imager, the approach could serve ground-based space situational awareness,
long-range ground-to-ground imaging and remote sensing. With faster SLMs, the approach could also work toward
correcting turbulence with the same pupil-plane modulator.

\begin{backmatter}

\bmsection{Funding}
This research is funded in part by the Army Research Office Modern Optics Program through DEPSCoR Grant.

\bmsection{Disclosures}
The authors declare no conflicts of interest.

\bmsection{Data Availability Statement}
Data underlying the results presented in this paper are not publicly available at this time but may be obtained from the authors upon request.

\end{backmatter}

\bibliography{root}

@misc{choi2025photonstarvedimagingturbulencediffraction,
      title={Photon-starved imaging through turbulence at the diffraction limit}, 
      author={Seungman Choi and Peter Menart and Andrew Schramka and Shubhankar Jape and Leif Bauer and In-Yong Park and Zubin Jacob},
      year={2025},
      eprint={2510.22806},
      archivePrefix={arXiv},
      primaryClass={physics.optics},
      howpublished = {arXiv:2510.22806},
      url={https://arxiv.org/abs/2510.22806}, 
}

@misc{zhang2025learningphasedistortionselective,
      title={Learning Phase Distortion with Selective State Space Models for Video Turbulence Mitigation}, 
      author={Xingguang Zhang and Nicholas Chimitt and Xijun Wang and Yu Yuan and Stanley H. Chan},
      year={2025},
      eprint={2504.02697},
      archivePrefix={arXiv},
      primaryClass={cs.CV},
      howpublished = {arXiv:2504.02697},
      url={https://arxiv.org/abs/2504.02697}, 
}

@article{neuws,
author = {Brandon Y. Feng  and Haiyun Guo  and Mingyang Xie  and Vivek Boominathan  and Manoj K. Sharma  and Ashok Veeraraghavan  and Christopher A. Metzler },
title = {NeuWS: Neural wavefront shaping for guidestar-free imaging through static and dynamic scattering media},
journal = {Science Advances},
volume = {9},
number = {26},
pages = {eadg4671},
year = {2023},
doi = {10.1126/sciadv.adg4671},
URL = {https://www.science.org/doi/abs/10.1126/sciadv.adg4671},
eprint = {https://www.science.org/doi/pdf/10.1126/sciadv.adg4671}}

@INPROCEEDINGS{refractive_index,
  author={Yasarla, Rajeev and Patel, Vishal M.},
  booktitle={2021 IEEE International Conference on Image Processing (ICIP)}, 
  title={Learning to Restore Images Degraded by Atmospheric Turbulence Using Uncertainty}, 
  year={2021},
  volume={},
  number={},
  pages={1694-1698},
  doi={10.1109/ICIP42928.2021.9506614}}

@misc{mao2020imagereconstructionstaticdynamic,
      title={Image Reconstruction of Static and Dynamic Scenes through Anisoplanatic Turbulence}, 
      author={Zhiyuan Mao and Nicholas Chimitt and Stanley Chan},
      year={2020},
      eprint={2009.00071},
      archivePrefix={arXiv},
      primaryClass={eess.IV},
      howpublished = {arXiv:2009.00071},
      url={https://arxiv.org/abs/2009.00071}, 
}

@article{gfws,
author = {Tomer Yeminy  and Ori Katz },
title = {Guidestar-free image-guided wavefront shaping},
journal = {Science Advances},
volume = {7},
number = {21},
pages = {eabf5364},
year = {2021},
doi = {10.1126/sciadv.abf5364},
URL = {https://www.science.org/doi/abs/10.1126/sciadv.abf5364},
eprint = {https://www.science.org/doi/pdf/10.1126/sciadv.abf5364}}

@article{mfbd,
author = {Timothy J. Schulz},
journal = {J. Opt. Soc. Am. A},
number = {5},
pages = {1064--1073},
publisher = {Optica Publishing Group},
title = {Multiframe blind deconvolution of astronomical images},
volume = {10},
month = {May},
year = {1993},
url = {https://opg.optica.org/josaa/abstract.cfm?URI=josaa-10-5-1064},
doi = {10.1364/JOSAA.10.001064},
}

@article{torchmfbd,
	author = {Asensio Ramos, A. and L{\"o}fdahl, M. G. and D{\'\i}az Baso, C. and Kuckein, C. and Esteban Pozuelo, S.},
	title = {torchmfbd: A flexible multi-object, multi-frame blind deconvolution code},
	DOI= "10.1051/0004-6361/202555530",
	url= "https://doi.org/10.1051/0004-6361/202555530",
	journal = {A\&A},
	year = 2025,
	volume = 703,
	pages = "A269",
}

@BOOK{AO,
       author = {{Hardy}, John W.},
        title = "{Adaptive Optics for Astronomical Telescopes}",
         year = 1998,
       adsurl = {https://ui.adsabs.harvard.edu/abs/1998aoat.book.....H}
}

@misc{rigaut2020multiconjugateadaptiveopticsastronomy,
      title={Multiconjugate Adaptive Optics for Astronomy}, 
      author={Francois Rigaut and Benoit Neichel},
      year={2020},
      eprint={2003.03097},
      archivePrefix={arXiv},
      primaryClass={astro-ph.IM},
      howpublished = {arXiv:2003.03097},
      url={https://arxiv.org/abs/2003.03097}, 
}

@article{kornilov2007,
   title={Combined {MASS--DIMM} instruments for atmospheric turbulence studies},
   volume={382},
   ISSN={0035-8711},
   url={http://dx.doi.org/10.1111/j.1365-2966.2007.12467.x},
   DOI={10.1111/j.1365-2966.2007.12467.x},
   number={3},
   journal={Monthly Notices of the Royal Astronomical Society},
   publisher={Oxford University Press (OUP)},
   author={Kornilov, V. and Tokovinin, A. and Shatsky, N. and Voziakova, O. and Potanin, S. and Safonov, B.},
   year={2007},
   month=Nov, pages={1268–1278} }

@article{fried1966,
  author  = {D. L. Fried},
  title   = {Optical resolution through a randomly inhomogeneous medium for very
             long and very short exposures},
  journal = {J. Opt. Soc. Am.},
  volume  = {56},
  number  = {10},
  pages   = {1372--1379},
  year    = {1966},
  doi     = {10.1364/JOSA.56.001372}
}

@article{noll1976,
  author  = {R. J. Noll},
  title   = {Zernike polynomials and atmospheric turbulence},
  journal = {J. Opt. Soc. Am.},
  volume  = {66},
  number  = {3},
  pages   = {207--211},
  year    = {1976},
  doi     = {10.1364/JOSA.66.000207}
}

@article{fried1982,
  author  = {D. L. Fried},
  title   = {Anisoplanatism in adaptive optics},
  journal = {J. Opt. Soc. Am.},
  volume  = {72},
  number  = {1},
  pages   = {52--61},
  year    = {1982},
  doi     = {10.1364/JOSA.72.000052}
}

@article{taylor1938,
  author  = {G. I. Taylor},
  title   = {The spectrum of turbulence},
  journal = {Proc. R. Soc. Lond. A},
  volume  = {164},
  number  = {919},
  pages   = {476--490},
  year    = {1938},
  doi     = {10.1098/rspa.1938.0032}
}

@book{andrews2005laser,
  author    = {L. C. Andrews and R. L. Phillips},
  title     = {Laser Beam Propagation through Random Media},
  edition   = {2nd},
  publisher = {SPIE Press},
  address   = {Bellingham, WA},
  year      = {2005},
  isbn      = {978-0-8194-5948-0},
  doi       = {10.1117/3.626196}
}

@article{lane1992screen,
  author  = {R. G. Lane and A. Glindemann and J. C. Dainty},
  title   = {Simulation of a {Kolmogorov} phase screen},
  journal = {Waves Random Media},
  volume  = {2},
  number  = {3},
  pages   = {209--224},
  year    = {1992},
  doi     = {10.1088/0959-7174/2/3/003}
}

@book{schmidt2010numerical,
  author    = {J. D. Schmidt},
  title     = {Numerical Simulation of Optical Wave Propagation with Examples in
               {MATLAB}},
  publisher = {SPIE Press},
  address   = {Bellingham, WA},
  year      = {2010},
  isbn      = {978-0-8194-8327-0},
  doi       = {10.1117/3.866274}
}

@article{gonsalves1982,
  author  = {R. A. Gonsalves},
  title   = {Phase retrieval and diversity in adaptive optics},
  journal = {Opt. Eng.},
  volume  = {21},
  number  = {5},
  pages   = {829--832},
  year    = {1982},
  doi     = {10.1117/12.7972989}
}

@article{paxman1992,
  author  = {R. G. Paxman and T. J. Schulz and J. R. Fienup},
  title   = {Joint estimation of object and aberrations by using phase diversity},
  journal = {J. Opt. Soc. Am. A},
  volume  = {9},
  number  = {7},
  pages   = {1072--1085},
  year    = {1992},
  doi     = {10.1364/JOSAA.9.001072}
}

@inproceedings{beckers1988,
  author    = {J. M. Beckers},
  title     = {Increasing the size of the isoplanatic patch with multiconjugate
               adaptive optics},
  booktitle = {Proc. ESO Conf. on Very Large Telescopes and their Instrumentation},
  volume    = {30},
  editor    = {M.-H. Ulrich},
  publisher = {ESO},
  pages     = {693--703},
  year      = {1988}
}

@article{vannoort2005,
  author  = {M. van Noort and L. {Rouppe van der Voort} and M. G. L{\"o}fdahl},
  title   = {Solar image restoration by use of multi-frame blind de-convolution
             with multiple objects and phase diversity},
  journal = {Sol. Phys.},
  volume  = {228},
  number  = {1--2},
  pages   = {191--215},
  year    = {2005},
  doi     = {10.1007/s11207-005-5782-z}
}

@article{nagy1998,
  author  = {J. G. Nagy and D. P. O'Leary},
  title   = {Restoring images degraded by spatially variant blur},
  journal = {SIAM J. Sci. Comput.},
  volume  = {19},
  number  = {4},
  pages   = {1063--1082},
  year    = {1998},
  doi     = {10.1137/S106482759528507X}
}

@inproceedings{hirsch2010,
  author    = {M. Hirsch and S. Sra and B. Sch{\"o}lkopf and S. Harmeling},
  title     = {Efficient filter flow for space-variant multiframe blind
               deconvolution},
  booktitle = {Proc. IEEE Conf. Computer Vision and Pattern Recognition (CVPR)},
  pages     = {607--614},
  year      = {2010},
  doi       = {10.1109/CVPR.2010.5540158}
}

@inproceedings{kingma2015adam,
  author        = {D. P. Kingma and J. Ba},
  title         = {Adam: a method for stochastic optimization},
  booktitle     = {3rd International Conference on Learning Representations (ICLR)},
  address       = {San Diego, CA},
  year          = {2015},
  eprint        = {1412.6980},
  archivePrefix = {arXiv}
}

@article{golub1973separable,
  author  = {G. H. Golub and V. Pereyra},
  title   = {The differentiation of pseudo-inverses and nonlinear least squares
             problems whose variables separate},
  journal = {SIAM J. Numer. Anal.},
  volume  = {10},
  number  = {2},
  pages   = {413--432},
  year    = {1973},
  doi     = {10.1137/0710036}
}

@article{rudin1992tv,
  author  = {L. I. Rudin and S. Osher and E. Fatemi},
  title   = {Nonlinear total variation based noise removal algorithms},
  journal = {Physica D},
  volume  = {60},
  number  = {1--4},
  pages   = {259--268},
  year    = {1992},
  doi     = {10.1016/0167-2789(92)90242-F}
}

@article{chao2016fisher,
  author  = {J. Chao and E. S. Ward and R. J. Ober},
  title   = {Fisher information theory for parameter estimation in single
             molecule microscopy: tutorial},
  journal = {J. Opt. Soc. Am. A},
  volume  = {33},
  number  = {7},
  pages   = {B36--B57},
  year    = {2016},
  doi     = {10.1364/JOSAA.33.000B36}
}

@article{lee1999crb,
  author  = {D. J. Lee and M. C. Roggemann and B. M. Welsh},
  title   = {Cram{\'e}r--{R}ao analysis of phase-diverse wave-front sensing},
  journal = {J. Opt. Soc. Am. A},
  volume  = {16},
  number  = {5},
  pages   = {1005--1015},
  year    = {1999},
  doi     = {10.1364/JOSAA.16.001005}
}

@article{rothenberg1971identification,
  author  = {T. J. Rothenberg},
  title   = {Identification in parametric models},
  journal = {Econometrica},
  volume  = {39},
  number  = {3},
  pages   = {577--591},
  year    = {1971},
  doi     = {10.2307/1913267}
}

@article{rockah1987array,
  author  = {Y. Rockah and P. Schultheiss},
  title   = {Array shape calibration using sources in unknown locations--{Part I}:
             Far-field sources},
  journal = {IEEE Trans. Acoust. Speech Signal Process.},
  volume  = {35},
  number  = {3},
  pages   = {286--299},
  year    = {1987},
  doi     = {10.1109/TASSP.1987.1165144}
}

@article{rocha2024mems,
  author  = {J. C. A. Rocha and T. Wright and U. G. B{\=u}tait{\.e} and
             J. Carpenter and G. S. D. Gordon and D. B. Phillips},
  title   = {Fast and light-efficient wavefront shaping with a {MEMS} phase-only
             light modulator},
  journal = {Opt. Express},
  volume  = {32},
  number  = {24},
  pages   = {43300},
  year    = {2024},
  doi     = {10.1364/OE.541857}
}

@misc{ximea2026camera,
  author       = {{XIMEA}},
  title        = {{CB024MG-GP-X8G3-SFF}: High speed camera with 2.4 {Mpix} monochrome
                  {Gpixel GSPRINT4502} sensor and {PCIe X8 Gen 3} interface},
  howpublished = {Product web page,
                  \url{https://www.ximea.com/products/pci-express/xib-64-high-speed-fast-industrial-cameras/gpixel-gsprint4502-pcie-mono-high-speed-camera}},
  note         = {Accessed 6 October 2026}
}

@article{labeyrie1970speckle,
  author  = {A. Labeyrie},
  title   = {Attainment of diffraction limited resolution in large telescopes by
             {Fourier} analysing speckle patterns in star images},
  journal = {Astron. Astrophys.},
  volume  = {6},
  pages   = {85--87},
  year    = {1970}
}

@article{roddier1981,
  author  = {F. Roddier},
  title   = {The effects of atmospheric turbulence in optical astronomy},
  journal = {Prog. Opt.},
  volume  = {19},
  pages   = {281--376},
  year    = {1981},
  doi     = {10.1016/S0079-6638(08)70204-X}
}

@article{schock2009tmt,
  author  = {M. Sch{\"o}ck and S. Els and R. Riddle and W. Skidmore and T. Travouillon and R. Blum and E. Bustos and G. Chanan and S. G. Djorgovski and P. Gillett and B. Gregory and J. Nelson and A. Ot{\'a}rola and J. Seguel and J. Vasquez and A. Walker and D. Walker and L. Wang},
  title   = {Thirty {Meter} {Telescope} site testing {I}: overview},
  journal = {Publ. Astron. Soc. Pac.},
  volume  = {121},
  number  = {878},
  pages   = {384--395},
  year    = {2009},
  doi     = {10.1086/599287}
}

\newpage
\appendix
\section*{Appendix}

\section{Simulation Details and Screen Statistics}
\label{app:sim}
\setcounter{equation}{0}
\renewcommand{\theequation}{\Alph{section}\arabic{equation}}
\providecommand{\theHequation}{}
\renewcommand{\theHequation}{\Alph{section}\arabic{equation}}%

We base our turbulence simulations using five assumptions of Section~\ref{sec:formation}. We assume one geometric thin layer with no scintillation and is seen from distant sources in monochromatic light. The block model of turbulence is a simplified approximation for per-pixel turbulence. The SLM is assumed to be ideal with known phases and instantaneous switching and instantaneous
exposures. The noise is Poisson, and the screen is frozen except under drift, a Taylor flow along $+x$ with no
boiling and no second layer. Measurement $k$ carries the code
\begin{equation}
  \theta_k(r) = \sum_{j=2}^{21} c_{kj} Z_j(r), \qquad c_{kj} \sim \mathcal{N}(0,1),
  \label{eq:code}
\end{equation}
where the $Z_j$ are the 20 Noll-indexed Zernike modes above piston, each of unit RMS over the pupil
$P$~\cite{noll1976}.

On the $G = 256$\,px grid the pupil has a 32\,px radius, giving 4\,px per $\lambda/D$, and the object fills a central
96\,px region of interest (ROI) with an 80\,px margin. The windows are $w_b(x, y) = q_i(y)\,q_j(x)$ with $b = 2i + j$,
$q_0 = 1 - q_1$ and
\begin{equation}
  \begin{aligned}
    q_1(x) &= \tfrac{1}{2} \bigl\{ 1 - \cos[\pi\, \xi(x)] \bigr\},\\
    \xi(x) &= \min\bigl\{ \max\bigl[ (x + \tfrac{1}{2} - 64)/128,\; 0 \bigr],\; 1 \bigr\},
  \end{aligned}
  \label{eq:window}
\end{equation}
for pixel index $x$, so each transition spans one block width, a taper of 1. With $D^2/(\lambda h) = 62.5$ the shifts
of Eq.~(\ref{eq:shift}) round to 16\,px.

Screens are drawn at the nominal $r_0 = 64/(\Dr)$\,px, with three levels of subharmonics, from the von K\'arm\'an
spectrum
\begin{equation}
  \mathcal{W}_\phi(u) = 0.023\, r_0^{-5/3} \exp\bigl[ -(u/u_m)^2 \bigr] \bigl( u^2 + L_0^{-2} \bigr)^{-11/6},
  \label{eq:psd}
\end{equation}
with $L_0 = 2560$\,px and $u_m = 5.92/(2\pi l_0)$ for an inner scale $l_0 = 0.001$\,px. With its aperture piston
removed, each screen is scaled by one factor to Noll's Kolmogorov value~\cite{fried1966,noll1976},
\begin{equation}
  \Phi \leftarrow \Phi \Bigl[ 1.03\,(\Dr)^{5/3} \Big/ \tfrac{1}{B} \textstyle\sum_b \mathrm{var}_P(\phi_b) \Bigr]^{1/2},
  \label{eq:nollscale}
\end{equation}
with $\phi_b$ the piston-removed crop at $s_b$, at the first measurement under drift. Screens marked actual skip this
rescaling. A screen is fixed by its seed and side, and each study used one realization per seed.
Fig.~\ref{fig:forward}(c) uses the 326\,px screen of seed 42. Seeds 42 and 43 gave 296\,px screens for
Fig.~\ref{fig:framecount} and 326\,px screens for Supplement~1, Sec.~S1, all scaled on the four $2\times2$ footprints
unless marked actual. Seeds 42 to 44 gave 340\,px screens for drift up to $\dtot = 4$ (Supplement~1, Sec.~S2) and
460\,px screens for Fig.~\ref{fig:timevarying}, one per seed serving every drift.

The per-realization rescaling of Eq.~(\ref{eq:nollscale}) is not standard and makes the screens stronger than their
label in the phase that blurs the image. At $L_0 = 40D$ the generator gives only 0.619 of Noll's value, the deficit
almost all in tip and tilt, so matching the total raises every spatial scale, by a factor that is itself random. Over
4000 draws at nominal 4, the piston-, tip- and tilt-removed aperture variance $\sigma^2_{\mathrm{PTT}}$ averages 2.1
times Noll's $\Delta_3 = 0.134\,(\Dr)^{5/3}$, against 1.0 unscaled. Averaged over the four block footprints,
$\sigma^2_{\mathrm{PTT}}$ defines the high-order measure of effective strength,
$4\,(\sigma^2_{\mathrm{PTT}}/\Delta_3)^{3/5}$, which spans 5.7 to 9.5 over the reconstruction screens at nominal 4.
The phase structure function at lags of 2, 4 and 8\,px gives 3.6 to 10.3, against 3.5 unscaled. Comparisons between
methods stay paired. Seeds with stronger effective turbulence give lower PSNR, as seed 43 does on the 460\,px screens
(9.5 in the high-order measure, against 6.2 and 5.9 for seeds 42 and 44). Effective strength is thus a likely, not
proven, contributor to the seed scatter.

A $2\times2$ partition is coarse against $\theta_0$. A field point at angular distance $\alpha$ from its block center
carries the mean-square phase error
\begin{equation}
  \sigma^2_{\mathrm{block}}(\alpha) = (\alpha/\theta_0)^{5/3},
  \label{eq:floor}
\end{equation}
which, averaged over a block $32\,\lambda/D$ wide, predicts 1.56\,rad$^2$ at $\Dr = 2$ and 4.94\,rad$^2$ at $\Dr = 4$.

The single-layer picture also needs the blur $\lambda/r_0$ to stay below $\theta_0$. In units of $\lambda/D$ the blur
is $\Dr$ and $\theta_0 = 0.314\,[D^2/(\lambda h)]/(\Dr)$, so with the nominal $r_0$ this holds only up to about
$\Dr = 4.4$. Beyond it lie the $\Dr = 6$, 8 and 10 conditions of Fig.~\ref{fig:framecount} and, by their effective
strengths, most nominal-4 screens. In a real atmosphere a profile $C_n^2(z)$ fixes $r_0$ through
$m_0 = \int C_n^2\, dz$ and $\theta_0$ through $m_{5/3} = \int C_n^2 z^{5/3}\, dz$, so a measured pair
$(r_0, \theta_0)$ determines the effective height
\begin{equation}
  \bar{h} = \bigl[ m_{5/3} / m_0 \bigr]^{3/5}
  \label{eq:hbar}
\end{equation}
exactly~\cite{fried1982}. One equivalent layer at $\bar h$ reproduces both, but not the footprint overlap that couples
neighboring blocks, and no multilayer case was simulated.

The coherence time $\tau_0 = 6.88^{-3/5} r_0/v = 0.314\,r_0/v$ of Section~\ref{sec:temporal} carries the coefficient of $\theta_0$,
since both mark the displacement, temporal or angular, at which the $5/3$-law phase structure function reaches
1\,rad$^2$. The screen travels $v\tau_0 = 0.314\,r_0$ per coherence time.

The registered PSNR is $-10\log_{10}$ of the mean-square error between the max-normalized estimate and truth over the
ROI, maximized over Fourier sub-pixel shifts on a 0.25\,px grid within $\pm 1.5$\,px of the cross-correlation peak.
The RMS phase error removes from each block's error the piston and tilt that maximize its phasor resultant over the
aperture eroded by 2\,px, refined by least squares. It then averages the wrapped residual's RMS over blocks. Removing
tilt per block also forgives differential tilt between blocks, which no object shift absorbs. Four references
accompany every reconstruction. The zero-phase reference, the phase error of a flat estimate, saturates near
$\pi/\sqrt{3} = 1.81$\,rad because of the wrap. Iteration~0, the estimator's starting point, is the closed-form object
at zero phase. The turbulent image is one uncoded long exposure with the full $1\times10^{9}$-photon budget through the
same windows, and the diffraction-limited reference, one noisy exposure through the unaberrated pupil at that budget,
scores 15.60--15.61\,dB.

\section{Estimator Implementation}
\label{app:est}
\setcounter{equation}{0}

At each spatial frequency, the closed-form object step of Section~\ref{sec:estimator} solves for the $B$ windowed pieces
$o_b = O w_b$ as independent unknowns, by least squares with ridge $\varepsilon K$ on the $K \times B$ matrix of the
optical transfer functions of the $H_{b,k}$. Least-squares re-windowing, $O = ( \sum_b w_b o_b ) / \sum_b w_b^2$, merges
the pieces into one object, and eight preconditioned projected Landweber passes built on this solve refine them. The
ridge is annealed geometrically from $10^{-1}$ to $10^{-5}$ over the run. Automatic differentiation through the whole
step gives the phase gradient.

Adam~\cite{kingma2015adam} with default moments runs 1500 iterations from one cold start at zero phase, at a constant
learning rate of 0.2. The rate was chosen on one realization, the seed-42 runs at $K = 64$, where at $\Dr = 10$ a rate
of 0.05 reached only 12.23\,dB against 17.17\,dB at 0.2. At $\Dr \le 4$ the total-variation weight is
$\mu = 0.01\,D_0$, with $D_0$ the initial data term, and a pixel pyramid runs the screen at 1/4, 1/2 and full
resolution over 20, 20 and 60\,\% of the run. Above $\Dr = 4$ the stronger weight $0.1\,D_0$, applied without the
pyramid, keeps the 0.2 rate stable, but the actual-strength series of Fig.~S5 of Supplement~1 uses the smaller
weight with the pyramid.

The reported object uses $\varepsilon = 10^{-3}$ with three passes, and the reported iterate has the lowest objective
at $\varepsilon = 10^{-5}$ among checkpoints every 25 iterations over the second half of the run. In the
continuous-drift study of Supplement~1, Sec.~S2, 46 of 60 coded reconstructions stopped at an interior iterate, and in
the other 14, which enter as measured, the loss was still falling when the budget ran out.

Replacing the closed-form step by a converged least-squares step, with a reporting ridge chosen against the true
object, raises the seed-mean registered PSNR on the matched $2\times2$ sky at $\Dr = 2$ from 18.49 to 19.11\,dB. The phase estimates are unchanged in quality as measured by PSNR, so on matched skies the
object step, not the phase, sets the PSNR. Every per-pixel reconstruction in the main text keeps the closed-form step.

\section{Fisher Information and Bounds}
\label{app:fisher}
\setcounter{equation}{0}
\renewcommand{\theequation}{\Alph{section}\arabic{equation}}
\providecommand{\theHequation}{}
\renewcommand{\theHequation}{\Alph{section}\arabic{equation}}

The identifiability analysis and the bounds of Figs.~\ref{fig:forward}(b) and~\ref{fig:rate}(b,~c) use the Fisher
information of the Poisson likelihood of Eq.~(\ref{eq:poisson}). For independent measurements with expected counts
$\bar n_k(r) = \gphot\, y_k(r)$, the information on a vector $\boldsymbol{\gamma}$ of real unknowns adds over
measurements and pixels~\cite{chao2016fisher},
\begin{equation}
  [\mathbf{F}]_{pq} = \sum_{k=1}^{K} \sum_{r} \frac{1}{\bar n_k(r)}\,
  \frac{\partial \bar n_k(r)}{\partial \gamma_p}\, \frac{\partial \bar n_k(r)}{\partial \gamma_q}.
  \label{eq:fisher}
\end{equation}
Any unbiased estimate then obeys
\begin{equation}
  \mathrm{Cov}(\hat{\boldsymbol{\gamma}}) \succeq \mathbf{F}^{-1},
  \label{eq:crb}
\end{equation}
where $\succeq$ means that the difference is positive semidefinite. The diagonal of $\mathbf{F}^{-1}$ is the CRB of
each unknown. Directions that leave every expected measurement unchanged, such as a phase piston, are gauges and are
removed before inversion. Any further null direction of $\mathbf{F}$ means the unknowns are not locally
identifiable~\cite{rothenberg1971identification}.

The rank tests use a 64\,px grid with a 16\,px pupil of $\Ap = 197$ pixels, and $N = 10^{6}$ photons from the test
scene, which fills the field. One 84\,px screen of seed 42 at $\Dr = 4$, scaled as in Eq.~(\ref{eq:nollscale}) on the
16 footprints of the $4\times4$ partition, serves every block count. The unknowns are the $\Ameta$ phase pixels of the
footprint union and, in the band-limited object model, the $\nband = 765$ real object coefficients inside the support
of the optical transfer function. Ranks are counted at a relative tolerance of $10^{-10}$ on the Jacobi-scaled
$\mathbf{F}$ after removing the gauges, one piston per connected footprint group and, at $B = 1$, the global tilt with
its compensating object shift. The CRB is that of the piston-, tip- and tilt-removed block phase, averaged over the
$B\Ap$ block pupil pixels with the object marginalized.

Under the band-limited object model, coded measurements are locally identifiable from $K = 2$ at $B = 1$, 4, 9 and
16, borderline at 16. For $B > 1$ the band limit is a prior rather than physics, because the windows move out-of-band
content into the band. Without it, two measurements do not identify the phase for $B = 4$, 9 and 16. Direct imaging
of a frozen screen with an unknown object is never identifiable, since its measurements are identical.
Identifiability is not precision. At $K = 2$ the mean phase CRB of this single realization is 4 to 6000 times the
true block-phase variance, and it first falls below that variance at $K = 3$, 5, 6 and 6 for $B = 1$, 4, 9 and 16.

The object-known bound of Fig.~\ref{fig:forward}(b) uses a reduced geometry, a 64\,px grid with a 20\,px pupil
of 317 pixels, $K = 10$ measurements, a central 20\,px object and one von K\'arm\'an screen scaled to Noll's value at
$\Dr = 20$. The anisoplanatic case has $2\times2$ hard quadrant masks and shifts of $\pm 6$\,px, 909 phase pixels in
all, and the coded case propagates the pupil field 50\,mm to the SLM and 50\,mm beyond by the angular-spectrum method
at 532\,nm. With the object known, the bound inverts only the phase block $\mathbf{F}_{\phi\phi}$ of $\mathbf{F}$. The
plotted quantity is the root of the per-pixel CRB averaged over the set $\mathcal{A}$ of the 317 on-axis aperture
pixels,
\begin{equation}
  \bar\sigma(N) = \Bigl[ \frac{1}{|\mathcal{A}|} \sum_{r \in \mathcal{A}} \bigl( \mathbf{F}_{\phi\phi}^{-1} \bigr)_{rr}
  \Bigr]^{1/2} \propto N^{-1/2}.
  \label{eq:sigmabar}
\end{equation}
Because $N\bar\sigma^2$ is constant along each curve, direct imaging needs 7.6 times as many photons as coded
measurements for equal $\bar\sigma$ in isoplanatic turbulence and 41 times as many in anisoplanatic turbulence.

The bound of Fig.~\ref{fig:rate}(b,~c) applies the same object-known reasoning to a screen in Taylor flow, in the
rank-test geometry with $2\times2$ blocks, shifts of $\pm 4$\,px, $K = 64$ and $1\times10^{9}$ photons per burst. The
screen moves as in Eq.~(\ref{eq:delta}) by exact Fourier shifts, with $r_0 = 16\,\mathrm{px}/(\Dr)$, so drift is
counted in nominal coherence times. On 352\,px ($\Dr = 2$) and 192\,px ($\Dr = 4$) screens of seed 42, the unknowns
are the pixels of the 25 screen rows that the footprints cross and the two wind components. A von K\'arm\'an prior at
the screens' nearly equal effective high-order strengths, $\Dr = 4.6$ and 5.1, adds its inverse covariance
$\mathbf{C}_\Phi^{-1}$ to $\mathbf{F}$, and the wind has no prior. Following the hybrid construction of
Ref.~\cite{rockah1987array}, the plotted root phase CRB is
\begin{equation}
  \bar\sigma_{\mathrm{tv}} = \Bigl\{ \frac{1}{K B \Ap}\,
  \operatorname{tr}\bigl[ \boldsymbol{\Gamma} \bigl( \mathbf{F} + \mathbf{C}_\Phi^{-1} \bigr)^{-1} \bigr] \Bigr\}^{1/2},
  \label{eq:tvbound}
\end{equation}
where $\boldsymbol{\Gamma}$ is the Gram matrix of the linear map from changes of screen and wind to the piston-,
tip- and tilt-removed phases of every block in every measurement, so a wind error counts as phase error. The 1.5\,kHz
SLM holds each pattern for one period from the first measurement, so a burst of $N < N^\ast$ coherence times holds
$\lfloor (K-1)N/N^\ast \rfloor + 1$ distinct patterns, the numbers printed in Fig.~\ref{fig:rate}(b,~c). The screens
differ mainly in travel per nominal coherence time, 2.5 and 1.3\,px, so the coded curves of panel (b) at $N$ coherence
times match those of panel (c) at $2N$ to about 1\,\%.

\clearpage
\section*{Supplement}
\setcounter{section}{0}
\renewcommand{\thesection}{S\arabic{section}}
\providecommand{\theHsection}{}
\renewcommand{\theHsection}{S\arabic{section}}
\setcounter{figure}{0}
\renewcommand{\thefigure}{S\arabic{figure}}
\providecommand{\theHfigure}{}
\renewcommand{\theHfigure}{S\arabic{figure}}

\section{Per-pixel Anisoplanatic Turbulence}\label{supp:blocks}

This section compares the block model of the main text with per-pixel anisoplanatism. In the per-pixel model every
object pixel has its own phase. The pixel at field angle $\alpha$ sees the layer through its own pupil footprint,
shifted as in Eq.~(\ref{eq:shift}) by the field angle times the distance between the layer and the aperture. The
per-pixel model is therefore much closer to real-world turbulence, and the block model is a simplified approximation
of it, in which the pixels of one block share the phase of the block center. The per-pixel model needs one PSF per
pixel and measurement, which makes it computationally expensive, so the main text uses the block model. We render measurements with per-pixel
anisoplanatism and estimate them with block models of varying pitch with respect to the isoplanatic angle.

\begin{figure*}[tp]
  \centering
  \includegraphics[width=1.0\textwidth]{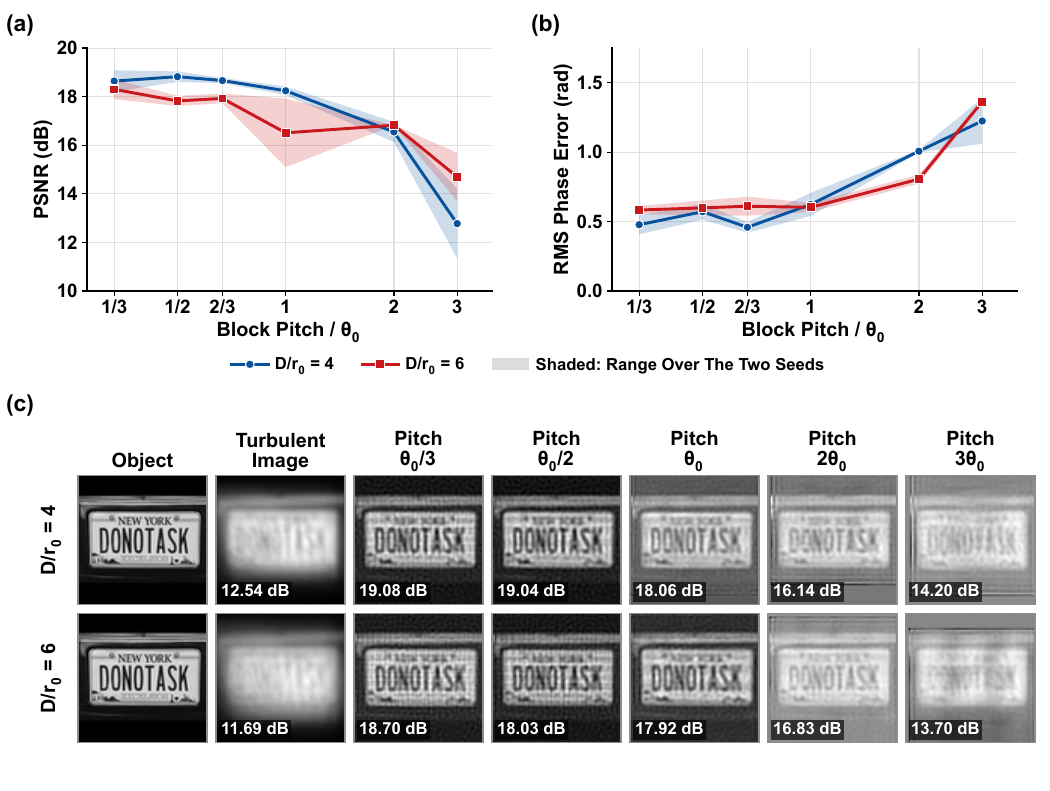}%
  \caption{Reconstruction of measurements with per-pixel anisoplanatism against the block pitch of the estimator, at
  actual $\Dr = 4$ and 6. The data are rendered with one PSF per object pixel from a single layer at 2~km (geometric
  optics, $K = 64$ coded measurements, $10^{9}$ photons), and the $\Dr = 6$ screens are the $\Dr = 4$
  screens scaled in strength. One phase screen is estimated on blocks from $\theta_0/3$ to $3\theta_0$ apart, with a
  conjugate-gradient object step.
  (a)~Registered PSNR: Lines give the mean over the two seeds and shaded bands their range.
  (b)~RMS error of the estimated phase over the pupil footprints of all object pixels, after removing one piston and
  one tilt.
  (c)~Reconstructions with the object and the turbulent image as estimator pitch size is increased ith respect to isopl;anatic angle.}
  \label{supp:fig:perpixel}
\end{figure*}

In Fig.~\ref{supp:fig:perpixel} the measurements are rendered with a separate
PSF for every object pixel, from one layer at $\Dr = 4$ and 6. The estimator still uses blocks as a
choice of how finely to sample the field rather than as a property of the data. It estimates one phase screen and
takes the PSF of each block from the footprint at the block center, with the block pitch varied from a third of the
isoplanatic angle $\theta_0$ to three times $\theta_0$. Because the data no longer follow the block model, the object
step is a conjugate-gradient solve rather than the closed-form step, and the estimator uses the smaller
total-variation weight with the pixel pyramid (Appendix~B).

Blocks spaced up to about one isoplanatic angle apart recover the object at both strengths. As the blocks become coarser than $\theta_0$, the phase error grows, because one block phase must
now describe pixels whose footprints differ, and at three times $\theta_0$ the reconstruction visibly degrades. The
block model is therefore a good approximation of per-pixel turbulence as long as its blocks are no wider than about one
isoplanatic angle.

\section{Time-Varying Turbulence}\label{supp:drift}

This section adds two time-varying turbulence studies to Fig.~\ref{fig:timevarying}. At  $\Dr = 6$ CA-BD stays close to its frozen-screen quality at every drift and leads both baselines in every
realization by a clear margin. At  $\Dr = 12$ it holds up for short bursts but loses quality as the burst spans
more coherence times, and at 16 coherence times its lead over the better baseline nearly vanishes in two of the three
realizations. In the crosswind example of Section~3 these strengths shorten the coherence time, so a 1.5\,kHz SLM can
give every measurement its own pattern only on bursts of about 16 coherence times at actual $\Dr = 6$ and about 33 at
actual $\Dr = 12$. The loss at $\Dr = 12$ is therefore consistent with the modulator rate being the limit in stronger
turbulence, but it does not prove it, since part of the loss may come from the estimator having few measurements per
epoch.

\begin{figure*}[t]
  \centering
  \includegraphics[width=1.0\textwidth]{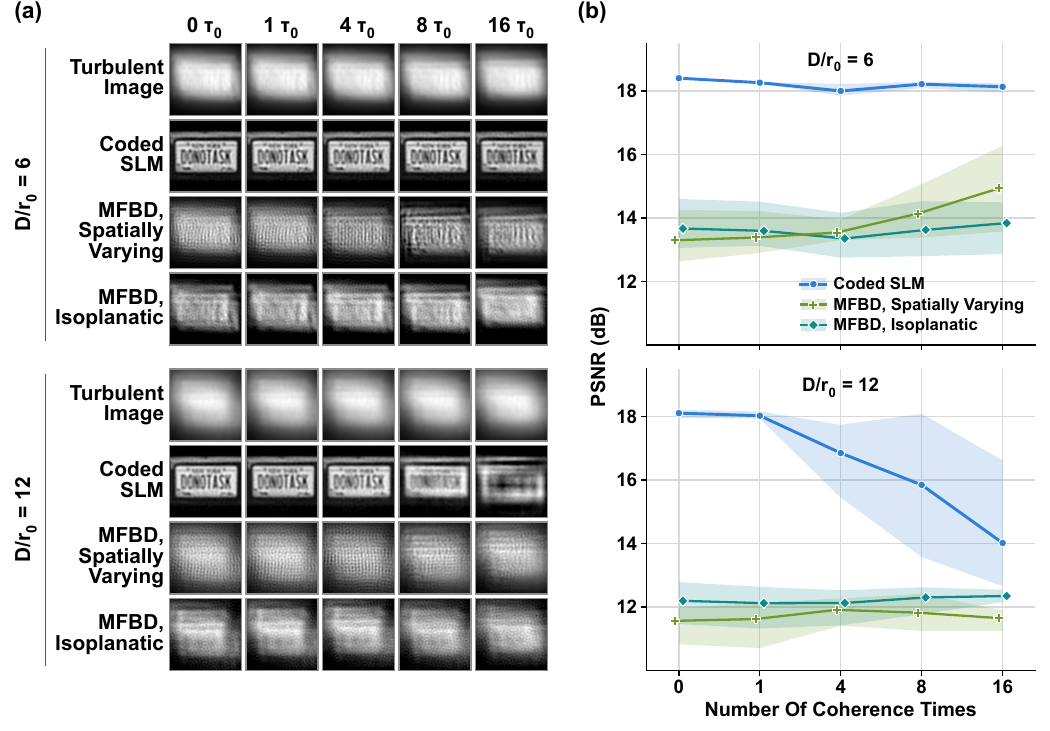}%
  \caption{Reconstruction through drifting turbulence at  $D/r_0 = 6$ (top) and 12 (bottom).
(a) Example reconstructions. Rows: turbulent image, CA-BD (Coded SLM) and the two MFBD
baselines; columns: number of coherence times spanned by the burst, from 0 (frozen screen) to 16.
(b) Registered PSNR versus number of coherence times; lines show the mean over three screen
realizations and bands their range. Screens are drawn as for Fig.~3 with $K = 64$ measurements and $10^9$ photons per burst. CA-BD uses one frozen epoch per
coherence time and the smaller total-variation weight with the pixel pyramid (Appendix~B). The
spatially varying MFBD uses a PSF dictionary trained only up to $D/r_0 \approx 12$.}
  \label{supp:fig:timevarying_actual}
\end{figure*}

We use the same estimation method of using one phase estimate per one epoch where one epoch corresponds to one $\tau_0$. We observer the as turbulence strength increases and as the amount of drift in the measurements increases, CA-BD loses performance due to two reasons - less information because of coded measurements due to the assumption of constant phase per $\tau_0$. This provides less information per $\tau_0$ estimate. We also notice that as turbulence strength increases, MFBD based methods fail to utilize diversity as diversity provided by turbulence becomes harder if turbulence change is severe. We emphasize that the specific region which has less drift across measurements, can recover severe turbulence because of added diversity. How far this regime extends depends on how quickly diversity can be added relative to
how fast the turbulence changes. We therefore expect our method to be limited by the refresh rate
of the SLM.

\end{document}